\documentclass[prb,showpacs,preprintnumbers,amsmath,amssymb,twocolumn]{revtex4-1}
\usepackage{graphicx}% Include figure files
\usepackage{dcolumn}% Align table columns on decimal point
\usepackage{bm}% bold math
\usepackage{amsmath}
\usepackage{amssymb}%this two are for type in greater or equal and less or equal sign
\usepackage{epstopdf}
\usepackage{gensymb}
\usepackage{array,multirow}
\usepackage{multirow}
\usepackage{dcolumn}
\usepackage{verbatim}
\usepackage{afterpage}
\usepackage{xfrac}
\usepackage{xcolor}
\usepackage[sort&compress]{natbib}
\usepackage[colorlinks=true,allcolors=blue]{hyperref}

\usepackage{latexsym}
\usepackage{graphicx}

\usepackage{subfigure}
\usepackage{epsfig}
\usepackage{float}

\begin{document}

%\templatetype{pnasresearcharticle} % Choose template 
% {pnasresearcharticle} = Template for a two-column research article
% {pnasmathematics} = Template for a one-column mathematics article
% {pnasinvited} = Template for a PNAS invited submission

\title{Correlated electron state in CeCu$_2$Si$_2$ controlled through Si to P substitution}

\author{Y. Lai$^{1,2}$, S. M. Saunders$^3$, D. Graf$^{1}$, A. Gallagher$^{1,2}$, K. W. Chen$^{1,2}$, F. Kametani$^{4}$, T. Besara,$^1$ T. Siegrist,$^{1,5}$ A. Shekhter$^{1}$, and R. E. Baumbach$^{1,2}$}
%\author{AUTHORS$^1$}
\affiliation{$^1$National High Magnetic Field Laboratory, Florida State University}
\affiliation{$^2$Department of Physics, Florida State University}
\affiliation{$^3$ Iowa State University and the Ames Laboratory}
\affiliation{$^4$Applied Superconductivity Center – Florida State University}
\affiliation{$^5$Department of Chemical and Biomedical Engineering, FAMU-FSU College of Engineering}
\date{\today}

% Keywords are not mandatory, but authors are strongly encouraged to provide them. If provided, please include two to five keywords, separated by the pipe symbol, e.g:

\begin{abstract}
CeCu$_2$Si$_2$ is an exemplary correlated electron metal that features two domes of unconventional superconductivity in its temperature-pressure phase diagram. The first dome surrounds an antiferromagnetic quantum critical point, whereas the more exotic second dome may span the termination point of a line of $f$-electron valence transitions. This behavior has received intense interest, but what has been missing are ways to access the high pressure behavior under milder conditions. Here we study Si $\rightarrow$ P chemical substitution, which compresses the unit cell volume but simultaneously weakens the hybridization between the $f$- and conduction electron states and encourages complex magnetism. At concentrations that show magnetism, applied pressure suppresses the magnetic ordering temperature and superconductivity is recovered for samples with low disorder. These results reveal that the electronic behavior in this system is controlled by a nontrivial combination of effects from unit cell volume and electronic shell filling. Guided by this topography we discuss prospects for uncovering a valence fluctuation quantum phase transition in the broader family of Ce-based ThCr$_2$Si$_2$-type materials through chemical substitution. 
\end{abstract}

%\begin{document}

% Optional adjustment to line up main text (after abstract) of first page with line numbers, when using both lineno and twocolumn options.
% You should only change this length when you've finalised the article contents.
%\verticaladjustment{-2pt}

\maketitle
%\thispagestyle{firststyle}
%\ifthenelse{\boolean{shortarticle}}{\ifthenelse{\boolean{singlecolumn}}{\abscontentformatted}{\abscontent}}{}

\section{Introduction}

Understanding magnetism, intermediate valence, unconventional superconductivity, and breakdown of Fermi liquid behavior in correlated electron materials is a persistent challenge.~\cite{mott,herring,Coleman01,Stewart01,Rosch07,Gegenwart08,Pfleiderer09} For Ce-based $f$-electron lattices, most discussion relies on a model where the Kondo and RKKY interactions compete with each other to determine the ground state behavior.~\cite{Doniach_77,kondo,ruderman,kasuya,yosida} While the Kondo interaction suppresses magnetism through screening of the $f$-electron moments by the conduction electrons, the RKKY interaction favors magnetic order by providing an exchange interaction between $f$-moments. Importantly, both interactions are mediated by the same wide band itinerant electrons which mostly originate on the $s$, $p$, and $d$ orbitals of the transition metal ions or ligand matrix. This scenario was described early-on in the Doniach phase diagram,~\cite{Doniach_77} and since then a multitude of compounds have been characterized in this way; noteworthy examples include Ce$T$In$_5$ ($T$ = Co, Rh, Ir)~\cite{thompson12,gofryk12} and Ce$T_2$$X_2$ ($T$ = transition metal and $X$ = Si, Ge).~\cite{palstra86,endstra} Although this picture has proved useful, efforts to develop it into a quantitative model have been challenged by the inherent complexity of the correlated electron problem; even the task of calculating the band structure for a cerium based metal remains difficult. Also missing is an understanding of how and where spin and charge instabilities play a role. To confront these issues, it is useful to re-examine the Kondo vs. RKKY scenario from a chemical perspective. 

It is interesting to focus on CeCu$_2$Si$_2$ because: (1) it exhibits a rich variety of behaviors including spin density wave magnetism, breakdown of the Fermi liquid, possible quantum criticality, and unconventional superconductivity~\cite{Steglich79,Knebel96,trovarelli97,yuan03,Yuan06,holmes07,Steglich12} and (2) it is located on the verge of maximum $d$-shell filling and has close-to the smallest unit cell volume of the naturally occuring 122 analogues - it is appealing to associate this with its electronic complexity. Of particular note is that it hosts a second superconducting dome at high pressure, which might be related to an $f$-electron valence change quantum phase transition.~\cite{kobayashi13,rueff,holmes07,miyake_07} The maximum superconducting transition temperature in this region is twice that of the low-pressure antiferromagnetic quantum phase transition, suggesting a strengthened superconducting pairing mechanism. This has inspired extensive work, but the required high pressure environment has slowed progress and even makes some techniques impossible (e.g., angle resolved photoemission spectroscopy). Thus, it would be useful to understand the underlying chemical/structural drivers and thereby uncover practical means of accessing related behavior under ambient conditions.

In this paper we study the chemical substitution series CeCu$_2$(Si$_{1-x}$P$_x$)$_2$ where even as the unit cell volume decreases, the parent compound superconductivity is rapidly suppressed and is replaced by magnetism that strengthens and becomes more anisotropic with increasing $x$. At the same time, the hybridization strength between the $f$- and conduction electron states is weakened. This conflicts with expectations from the Doniach picture, where in the vicinity of a quantum critical point a decreasing unit cell volume is expected to strengthen the hybridization strength and suppress magnetism.~\cite{Doniach_77} To understand this, we consider a simplified phase map that is parameterized by the axes of unit cell volume and electronic shell filling. We suggest that Si $\rightarrow$ P substutution explores a nontrivial vector between these axes, where the impact of shell filling outweighs that of decreasing volume in this particular case. To probe the influence of further unit cell contraction without additional shell filling, measurements under applied pressure were carried out. Here, the magnetism is suppressed and superconductivity is recovered at a quantum phase transition, for lightly substituted samples. Based on these results, we discuss chemical strategies to induce behavior similar to that of high pressure phase CeCu$_2$Si$_2$ within the broader family of Ce-based ThCr$_2$Si$_2$-type materials.

%We further find that within this topography, the magnetic and non-magnetic members of the Ce-based ThCr$_2$Si$_2$-type materials are well separated and the crossover between these regions is populated by those examples that exhibit non-Fermi liquid behavior and unconventional superconductivity, providing a useful way to organize their behavior.  

\section{Experimental Methods}

Single crystals of CeCu$_2$Si$_{2-x}$P$_x$ were grown from elements with purities\(>\) 99.9\% in a molten flux of Cu and Si. The reaction ampoules were prepared by loading the elements in the ratio Ce:Cu:Si:P $;$ 1:28:(11.44$-$$x$):$x$ into a 2 mL alumina crucible for each of the different nominal dopings of P. The crucibles were sealed under vacuum in quartz ampoules and heated to $600\,^{\circ}\mathrm{C}$ at a rate of $50\,^{\circ}\mathrm{C}$/hour, held at $600\,^{\circ}\mathrm{C}$ for 6 hours, heated to $1185\,^{\circ}\mathrm{C}$ at a rate of $50\,^{\circ}\mathrm{C}$/hour, kept at $1185\,^{\circ}\mathrm{C}$ for 12 hours, and then cooled at a rate of $2\,^{\circ}\mathrm{C}$/hour to $940\,^{\circ}\mathrm{C}$. At this temperature, the remaining flux was separated from the crystals by centrifuging. Single-crystal platelets with typical dimensions of several millimeters on a side and several millimeters in thickness were collected.

The crystal structure and chemical composition were verified by single-crystal x-ray-diffraction (XRD) and energy dispersive spectrometer (EDS) analysis. A comparison between the nominal $x_{\rm{nom}}$ and actual $x_{\rm{act}}$ values is shown in Fig.~\ref{figS2}a. Throughout the manuscript, $x$ refers to the measured value. Magnetization $M$($T,H$) measurements were carried out for single crystals at temperatures $T$ $=$ 1.8 $-$ 300 K under an applied magnetic field of $H$ $=$ 5 kOe for $H$ applied both parallel ($\parallel$) and perpendicular ($\perp$) to the $c$ axis using a Quantum Design VSM Magnetic Property Measurement System. Electrical resistivity $\rho$ measurements for temperatures $T$ $=$ 0.5 $-$ 300 K and magnetic fields $H$ $=$ 0 $-$ 9 T were performed in a four-wire configuration and the heat capacity $C$ was measured for $T$ $=$ 0.39 $-$ 20 K using a Quantum Design Physical Property Measurement System. Measurements under applied pressure were performed using a piston cylinder pressure cell with the pressure transmitting medium Daphne 7474 oil. The pressure is determined by the shift in ruby flourescence peaks and are the values determined below $T$ $=$ 10 K. These measurements were performed at the National High Magnetic Field Laboratory DC field User facility using standard He3 cryostats.

\section{Results}

\begin{figure}%[tbhp]
\centering
\includegraphics[width=1\linewidth]{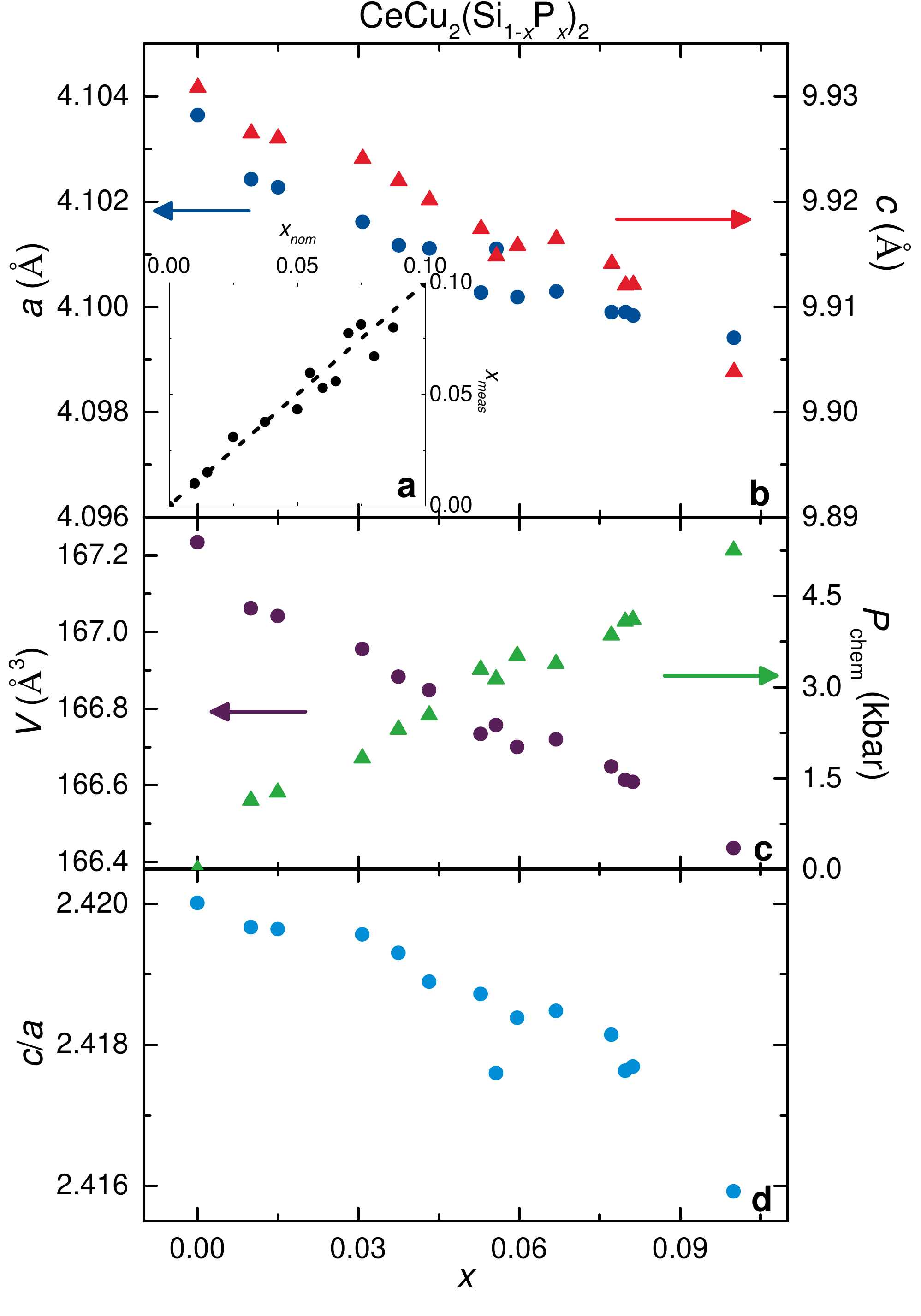}
\caption{(a) Comparison between the measured phosphorous concentration $x_{\rm{meas}}$ and the nominal concentration $x_{\rm{nom}}$, where $x_{\rm{meas}}$ was determined using energy dispersive spectrometer analysis. Throughout the manuscript we use $x_{\rm{meas}}$ $=$ $x$ unless otherwise specified. (b) The lattice constants, $a(x)$ (left axis) and $c(x)$ (right axis). (c) the unit cell volume $V(x)$ (left axis) and chemical pressure $P_{\rm{ch}}$(right axis), calculated using the Burch-Murnaghan equation, where $B_0$ $=$ 110 GPa.~\cite{spain} (d) the ratio $c/a$ vs. $x$.}
\label{figS2}
\end{figure}

Single crystal x-ray diffraction measurements show that Si $\rightarrow$ P substitution up to $x$ $\approx$ 0.1 causes the lattice constants ($a$ and $c$), the unit cell volume ($V$), and the ratio $c/a$ to decrease linearly (Fig.~\ref{figS2}). This is consistent with Vegard's law, where the monotonous lattice contraction is due to the smaller size of P by comparison to Si. From this we infer that the Ce valence remains roughly constant with increasing $x$. These changes result in a positive chemical pressure which is estimated to be near $P_{\rm{ch}}$ $=$ $B_{\rm{0}}$ln($V_0$/V) $\approx$ 4.8 kbar for $x$ $\approx$ 0.1, using the Birch-Murnaghan equation of state. Here we approximate the bulk modulus as $B_{\rm{0}}$ = 110 GPa as previously reported for stoichiometric CeCu$_2$Si$_2$.~\cite{spain}

\begin{figure}%[tbhp]
\centering
\includegraphics[width=1\linewidth]{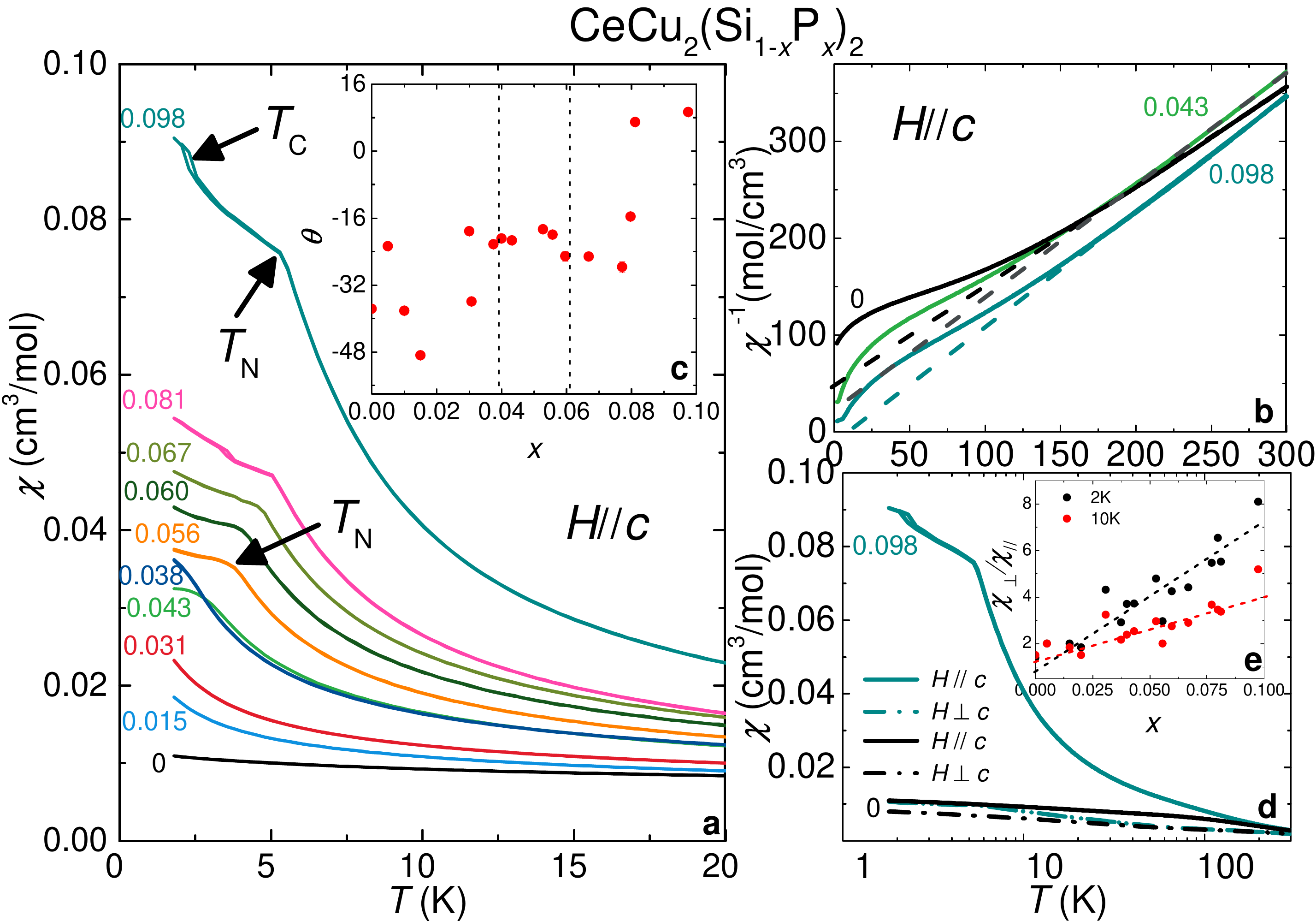}
\caption{(a) Magnetic susceptibility $\chi$ $=$ $M/H$ collected in a magnetic field $H$ $=$ 5 kOe applied $\parallel$ to the $c$-axis of CeCu$_2$(Si$_{1-x}$P$_x$)$_2$ for $x$ $=$ 0 - 0.1. (b) The inverse magnetic susceptibility $\chi$$^{-1}$ vs $T$ for $H$ $\parallel$ $c$ at concentrations $x$ $=$ 0, 0.043, and 0.098. The dotted lines are Curie-Weiss fits to the data using the expression $\chi(T)$ $=$ $C$/($T$-$\Theta$). (c) The Curie-Weiss temperature $\theta$ extracted from fits to $\chi(T)$ vs. $x$. (d) Magnetic susceptibility $\chi$ vs. temperature $T$ for CeCu$_2$(Si$_{1-x}$P$_x$)$_2$ at select concentrations $x$ $=$ 0 and 0.098 for magnetic fields $H$ applied parallel $\parallel$ (solid line) and perpendicular $\perp$ (dotted line) to the $c$-axis. (e) The ratio of the magnetic susceptibilities for $H$ $\parallel$ and $\perp$ to the $c$-axis at $T$ $=$ 2 and 10 K vs. $x$.}
\label{figS1}
\end{figure}

The $f$-electron behavior is uncovered in more detail by considering the magnetic susceptibility ($\chi$ = $M/H$) temperature dependence (Fig.~\ref{figS1}). Curie-Weiss fits to $\chi(T)$ for $H$ $\parallel$ c on the range 175 $<$ $T$ $<$ 300 K show that the effective magnetic moment $\mu_{\rm{eff}}$ $\approx$ 2.5 - 2.6 $\mu_{\rm{B}}$/Ce remains nearly constant and the Curie-Weiss temperature $\Theta$ is negative up to $x$ $\approx$ 0.08 where it becomes positive (Fig.~\ref{figS1}b). The magnetism becomes more anisotropic with increasing $x$, as evidenced by the strengthening $\chi$($T$) for $H$ $\parallel$ $c$ that yields a nearly tenfold increase in the low temperature magnetic anisotropy between $x$ = 0 - 0.1 (Figs.~\ref{figS1}c,d). This shows that while phosphorous substitution controls the magnetocrystalline anisotropy, it does not appreciably change the cerium valence. A complex evolution of the magnetic ground state is also observed. Up to $x$ $\approx$ 0.03 - 0.04 there is no obvious magnetic ordering seen in $\chi(T)$, but above this concentration there is a pronounced kink (labeled $T_{\rm{N}}$) which reduces $\chi$, consistent with antiferromagnetic ordering. $T_{\rm{N}}$ subsequently sharpens and moves to higher $T$ with increasing $x$. Near $x$ $\approx$ 0.06 - 0.07 a second small hysteretic increase in $\chi$ appears at $T_{\rm{C}}$ $<$ $T_{\rm{N}}$, revealing an additional spin reconfiguration.  

\begin{figure}%[tbhp]
\centering
\includegraphics[width=1\linewidth]{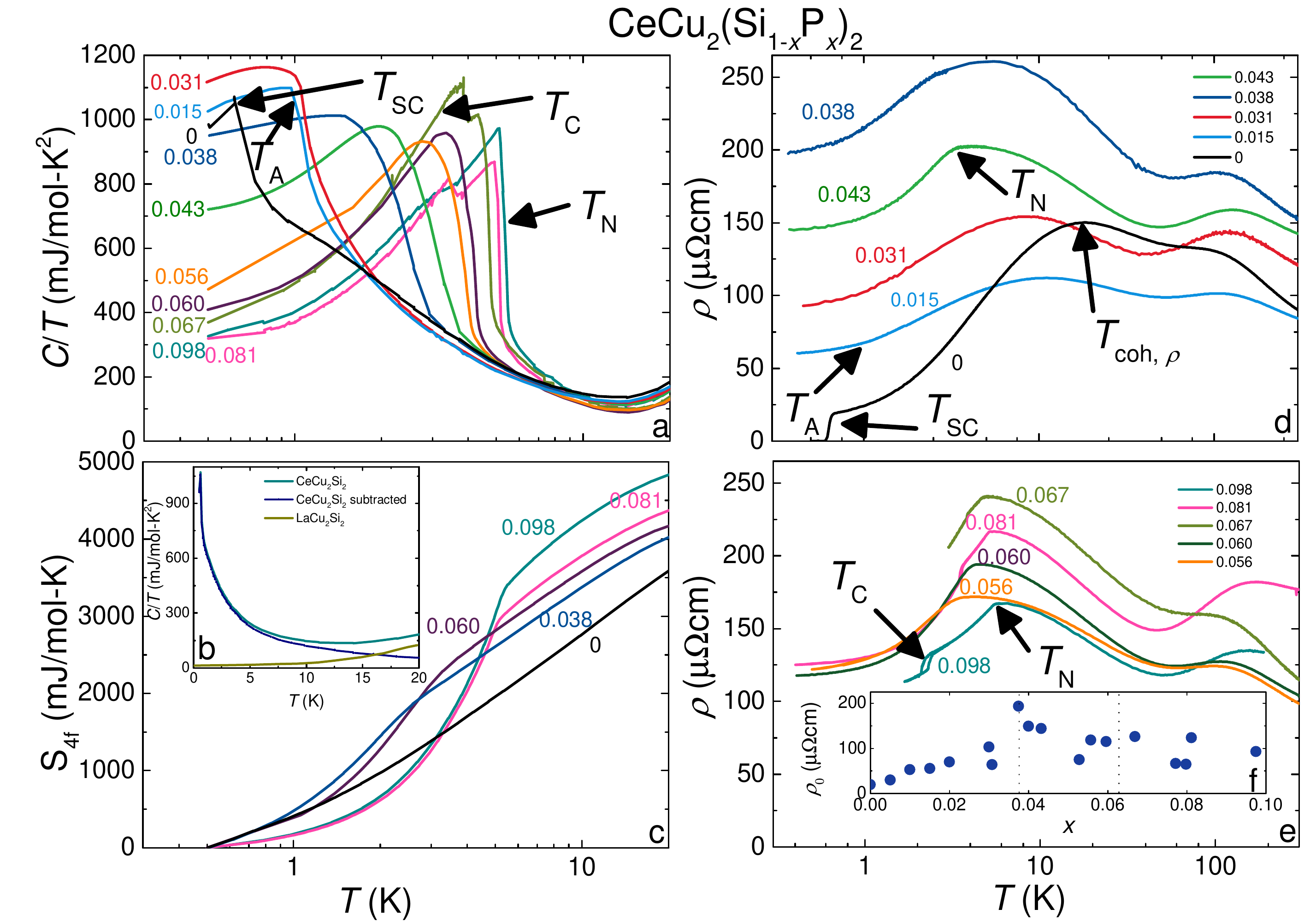}
\caption{(a) Heat capacity $C$ divided by temperature $T$ vs. $T$ of CeCu$_2$(Si$_{1-x}$P$_x$)$_2$ for $x$ $=$ 0 - 0.1. (b) $C/T$ vs. $T$ for CeCu$_2$Si$_2$ and LaCu$_2$Si$_2$. Also shown is the 4$f$ contribution to the heat capacity of CeCu$_2$Si$_2$ (blue circles) $C_{4f}$/$T$, which was acquired by subtracting the La contribution from that of the Ce compound. (c) The 4$f$ contribution to the entropy $S_{4f}$ vs. $T$ for CeCu$_2$(Si$_{1-x}$P$_x$)$_2$. $S_{4f}$ was acquired as described in the text. (d) The electrical resistivity $\rho$ vs. $T$ for 0 $<$ $x$ $<$ 0.043. (e) $\rho$ vs. $T$ for 0.056 $<$ $x$ $<$ 0.098. (f) Residual resistivity $\rho_0$ vs. $x$. }
\label{figS3}
\end{figure}

The temperature dependence of the heat capacity divided by temperature $C/T$ and the electrical resistivity are shown in Fig.~\ref{figS3}. Data for $x$ $=$ 0 are consistent with earlier results for `S-phase' CeCu$_2$Si$_2$, where $C/T$ increases to a large value at low temperatures due to the large charge carrier quasiparticle mass. There is a sharp feature near $T_{\rm{SC}}$ $\approx$ 0.6 K in $C/T$ and a drop to zero resistivity at the onset of superconductivity.~\cite{Steglich79} The superconductivity is rapidly destroyed with $x$ and is replaced by a new ordered phase that is evidenced by a weak kink in $\rho$ and a lambda-anomaly in $C/T$ at $T_{\rm{A}}$. This is similar to the `A-phase' antiferromagnetism that is seen in self-doped CeCu$_2$Si$_2$ and CeCu$_2$(S$_{1-x}$Ge$_x$)$_2$~\cite{Knebel96,trovarelli97}. Near $x$ $\approx$ 0.03 $-$ 0.04, the kink in $\rho$($T$) is replaced by an abrupt reduction in the resistivity at $T_{\rm{N}}$ and a broadened lambda-like anomaly in $C/T$. For $x$ $\gtrsim$ 0.06 - 0.07 the additional phase transition at $T_{\rm{C}}$ is also seen in $\rho$($T$) and $C/T$, showing that it occurs in the bulk. From $\rho(T)$ it is also possible to identify a broad low temperature hump that precedes the low temperature ordering. This feature was earlier described as being related to the Kondo coherence temperature $T_{\rm{coh},\rho}$ $\approx$ 18 K for $x$ $=$0. $T_{\rm{coh},\rho}$ decreases with $x$ and becomes comparable to the magnetic ordering temperature near $x$ $\approx$ 0.04, which is roughly where the magnetic ordering changes its character.

The 4$f$-contribution to the entropy $S_{4f}$ is shown in Fig.~\ref{figS3}c, which was acquired by subtracting $C/T$ for LaCu$_2$Si$_2$ from the chemically substituted specimens and subsequently integrating $C_{4f}$/$T$ from $T$ $=$ 400 mK (Fig.~\ref{figS3}b). While this procedure underestimates $S_{4f}$, it provides a systematic way to compare between concentrations. From this, it is seen that the amount of entropy that is recovered at the magnetic ordering temperature increases with increasing $x$, reaching slightly more than 0.5$R$ln2 at maximum $x$. $S_{4f}$ $=$ $R$ln2 is expected for trivalent cerium when crystal electric field splitting of the Hund's rule multiplet results in a doublet ground state, but this value is reduced when the $f$-electron moment is compensated through the Kondo interaction. Thus, the increasing value of $S_{4f}$($x$) at the ordering temperature suggests weakening hybridization without a complete removal of Kondo screening at $x$ $\approx$ 0.1. 

%\begin{figure}%[tbhp]
%\centering
%\includegraphics[width=1\linewidth]{FigSR.pdf}
%\caption{(a) The electrical resistivity $\rho$ vs. $T$ for 0 $<$ $x$ $<$ 0.043. (b) $\rho$ vs. $T$ for 0.056 $<$ $x$ $<$ 0.098. (c) Residual %resistivity $\rho_0$ vs. $x$. }
%\label{figSR}
%\end{figure}

\begin{figure}%[tbhp]
\centering
\includegraphics[width=1\linewidth]{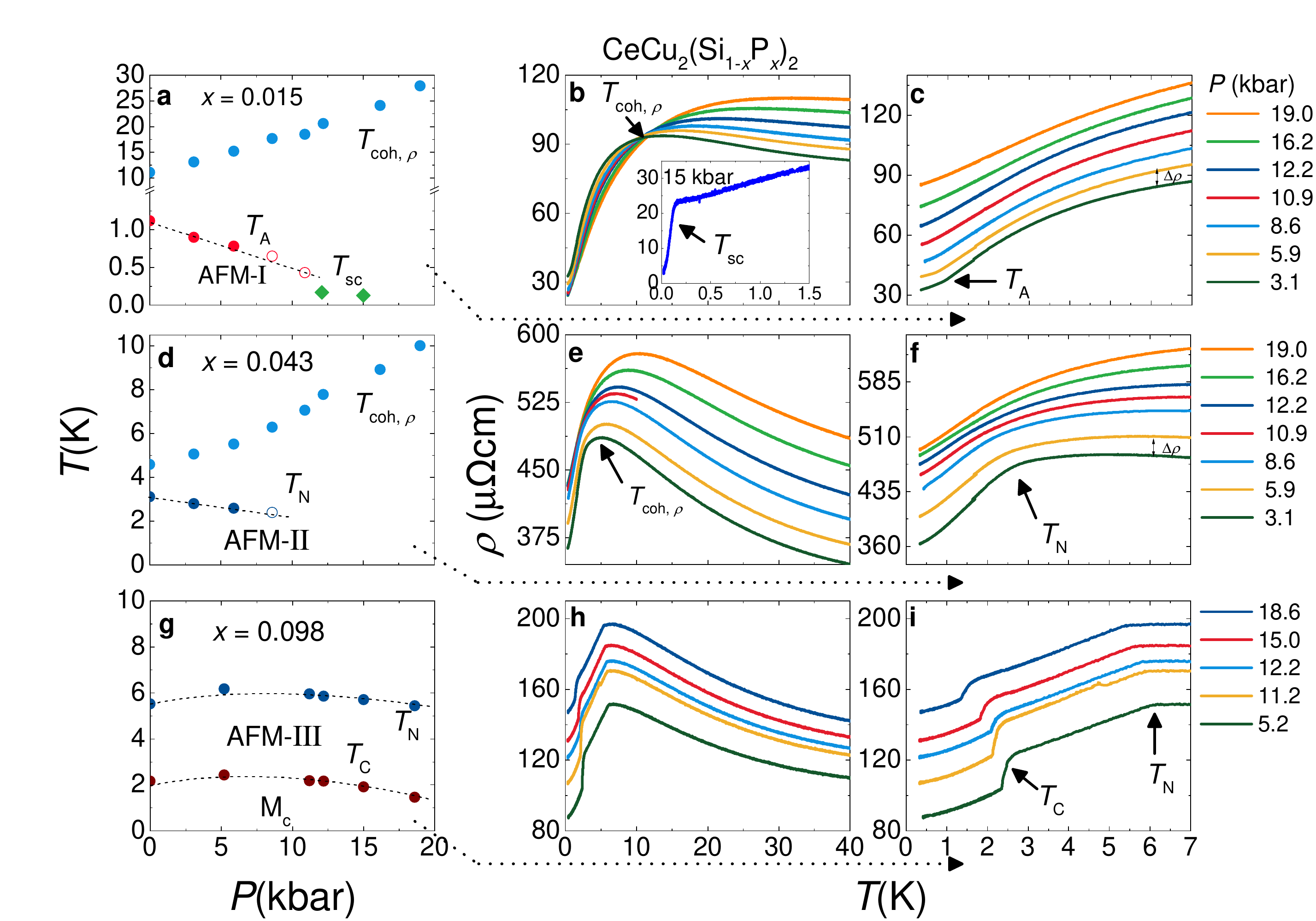}
\caption{Summary of electrical resistivity $\rho$ vs. temperature $T$ measurements under applied pressure $P$ for CeCu$_2$(Si$_{1-x}$P$_x$)$_2$ at select concentrations $x$ = 0.015, 0.043, and 0.098. For clarity, the curves for $x$ $=$ 0.015 and 0.043 are offset by constant values $\Delta$$\rho$ $=$ 10 $\mu$$\Omega$cm. (a) Temperature $T$ - pressure $P$ phase diagram for $x$ = 0.015 for $P$ $<$ 20 kbar showing the suppression of $T_{\rm{A}}$, the enhancement of $T_{\rm{coh},\rho}$, and the appearance of superconductivity at the extrapolated quantum phase transition. Open circles indicate ambiguity in defining $T_{\rm{A}}$. (b) $\rho$($T$) for $x$ = 0.015 for $P$ $<$ 20 kbar at 0 $<$ $T$ $<$ 40 K where the Kondo coherence temperature $T_{\rm{coh},\rho}$ appears as a broad hump. (c) $\rho$($T$) for $x$ = 0.015 for pressures $P$ $<$ 20 kbar and 0 $<$ $T$ $<$ 7 K. (d) $T-P$ phase diagram for $x$ = 0.043 for $P$ $<$ 20 kbar. (e) $\rho$($T$) for $x$ = 0.043 for $P$ $<$ 20 kbar at 0 $<$ $T$ $<$ 40 K. (f) $\rho$($T$) for $x$ = 0.043 for pressures $P$ $<$ 20 kbar at 0 $<$ $T$ $<$ 7 K. (g) $T-P$ phase diagram for $x$ = 0.098 for several pressures $P$ $<$ 20 kbar. (h) $\rho$($T$) for $x$ = 0.098 for several pressures $P$ and 0 $<$ $T$ $<$ 40 K. (i) $\rho$($T$) for $x$ = 0.098 at $P$ $<$ 20 kbar and 0 $<$ $T$ $<$ 7 K.}
\label{fig3}
\end{figure}

In order to further examine the surrounding phase diagram, we performed measurements of the electrical resistivity under hydrostatic pressures for select concentrations ($x$ $=$ 0.015, 0.043, 0.098) spanning $T-x$ phase diagram (Fig.~\ref{fig3}). For $x$ = 0.015 and 0.045, both $T_{\rm{A}}$ and $T_{\rm{N}}$ are suppressed with initial slopes $\partial$$T_{\rm{A}}$/$\partial$$P$ = 0.06 K/kbar and $\partial$$T_{\rm{N}}$/$\partial$$P$ = 0.07 K/kbar and are extrapolated to approach zero temperature near $P_{\rm{c}}$ $\approx$ 12 $-$ 16 and 18 $-$ 22 kbar, respectively. Near $P$ $=$ 15 kbar for $x$ $=$ 0.015, superconductivity is recovered with an onset temperature $T_{\rm{SC}}$ $\approx$ 150 mK and the normal state electrical resistivity follows the temperature dependence $\rho$($T$) = $\rho_0$ + $A$$T^n$ where $n$ $\approx$ 1.16 (Fig.~\ref{fig3}b), indicating a departure from Fermi liquid behavior. The superconductivity is noticeably robust against disorder, which is revealed here in the large residual resistivity. We further note that for both of these concentrations, $T_{\rm{coh},\rho}$ increases with increasing $P$, as is expected upon strengthening the hybridization between the $f$ - and conduction electrons. For $x$ = 0.098, both magnetic ordering temperatures $T_{\rm{N}}$ and $T_{\rm{C}}$ are suppressed by pressure with slopes $\partial$$T_{\rm{N}}$/$\partial$$P$ = 0.06 K/kbar and $\partial$$T_{\rm{C}}$/$\partial$$P$ $=$ 0.1 K/kbar, where the extrapolated critical pressures are above 20 kbar. 

\section{Discussion}
\begin{figure}%[tbhp]
\centering
\includegraphics[width=1\linewidth]{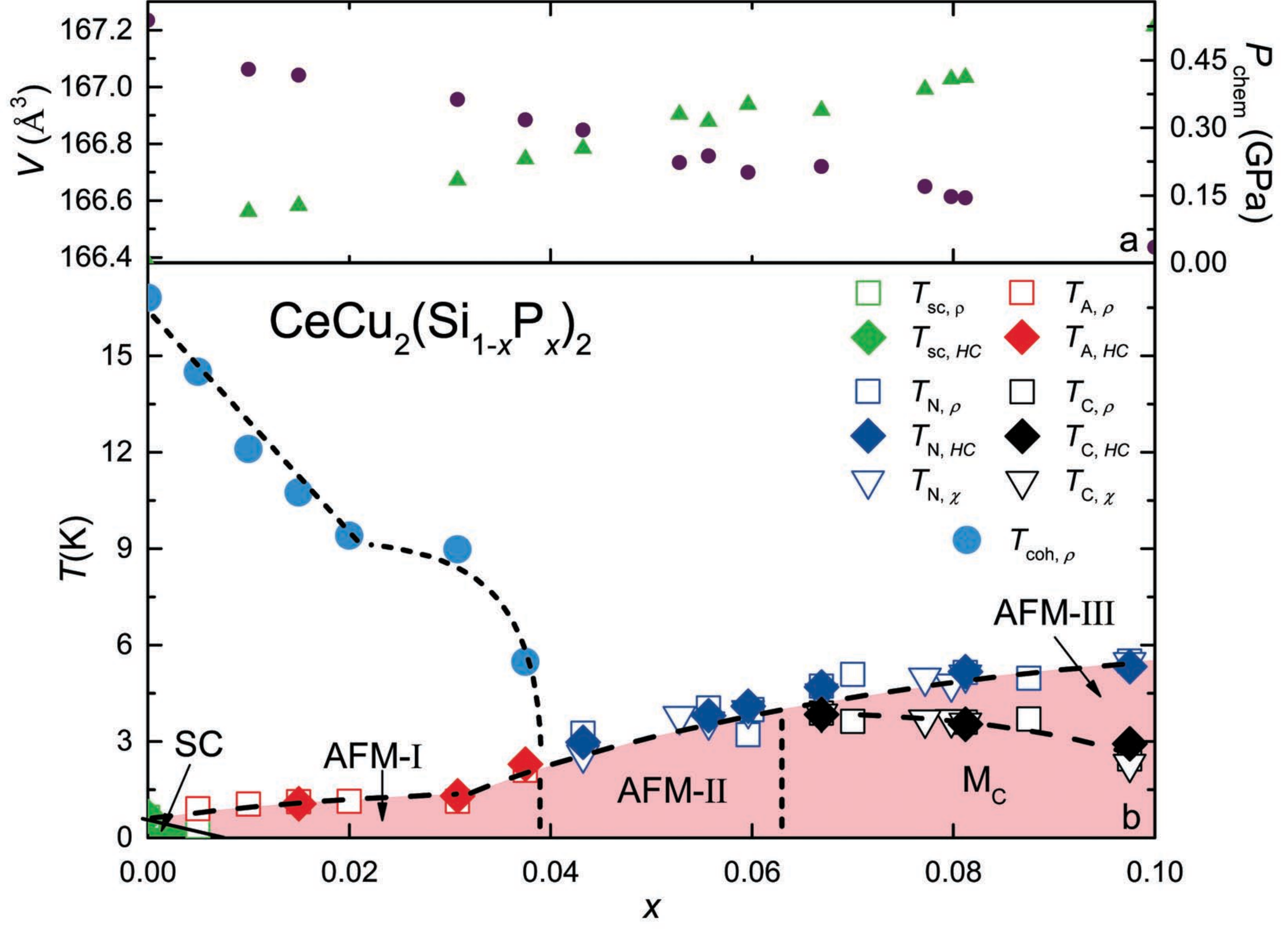}
\caption{Summary of thermodynamic and electrical transport results for CeCu$_2$(Si$_{1-x}$P$_x$)$_2$ at concentrations $x$ = 0 $-$ 0.1. (a) Unit cell volume $V$ (left axis) and chemical pressure $P_{\rm{chem}}$ vs. $x$, where $x$ is the measured value. $P_{\rm{chem}}$ was obtained using the Birch-Murnaghan equation of state. (b) Temperature $T$ - phosphorous concentration $x$ phase diagram for CeCu$_2$(Si$_{1-x}$P$_x$)$_2$ constructed from magnetic susceptibility $\chi$, heat capacity $C$, and electrical resistivity $\rho$ vs. $T$ measurements. The magnetic ordering temperatures $T_{\rm{A}}$, $T_{\rm{N}}$, and $T_{\rm{C}}$ and the Kondo coherence temperature $T_{\rm{coh,\rho}}$ are defined as described in the text.}
\label{fig2}
\end{figure}

Taken together, these measurements reveal a phase diagram (Fig.~\ref{fig2}) that is notably similar to that of CeCu$_2$(S$_{1-x}$Ge$_x$)$_2$,~\cite{Knebel96,trovarelli97} where Si $\rightarrow$ P (or Ge) substitution rapidly suppresses the parent compound superconductivity, induces complex magnetism, and weakens the hybridization between the $f$- and conduction electrons (e.g., as evidenced by a decreasing Kondo coherence temperature $T_{\rm{coh},\rho}$ and increasing $S_{4f}$ at the magnetic ordering temperature).  On its surface, this result is unexpected because Si $\rightarrow$ P and Si $\rightarrow$ Ge substitution result in volume contraction and expansion, respectively: i.e., they should have opposite affects. We understand this contradiction by considering that the hybridization strength is governed both by the unit cell volume and electronic shell filling, where non-isoelectronic chemical substitution explores a non-trivial vector amongst them. In our study, shell filling is the dominant term and has the effect of weakening the hybridization. 

\begin{figure}[t]
    \begin{center}
        \includegraphics[width=1\linewidth]{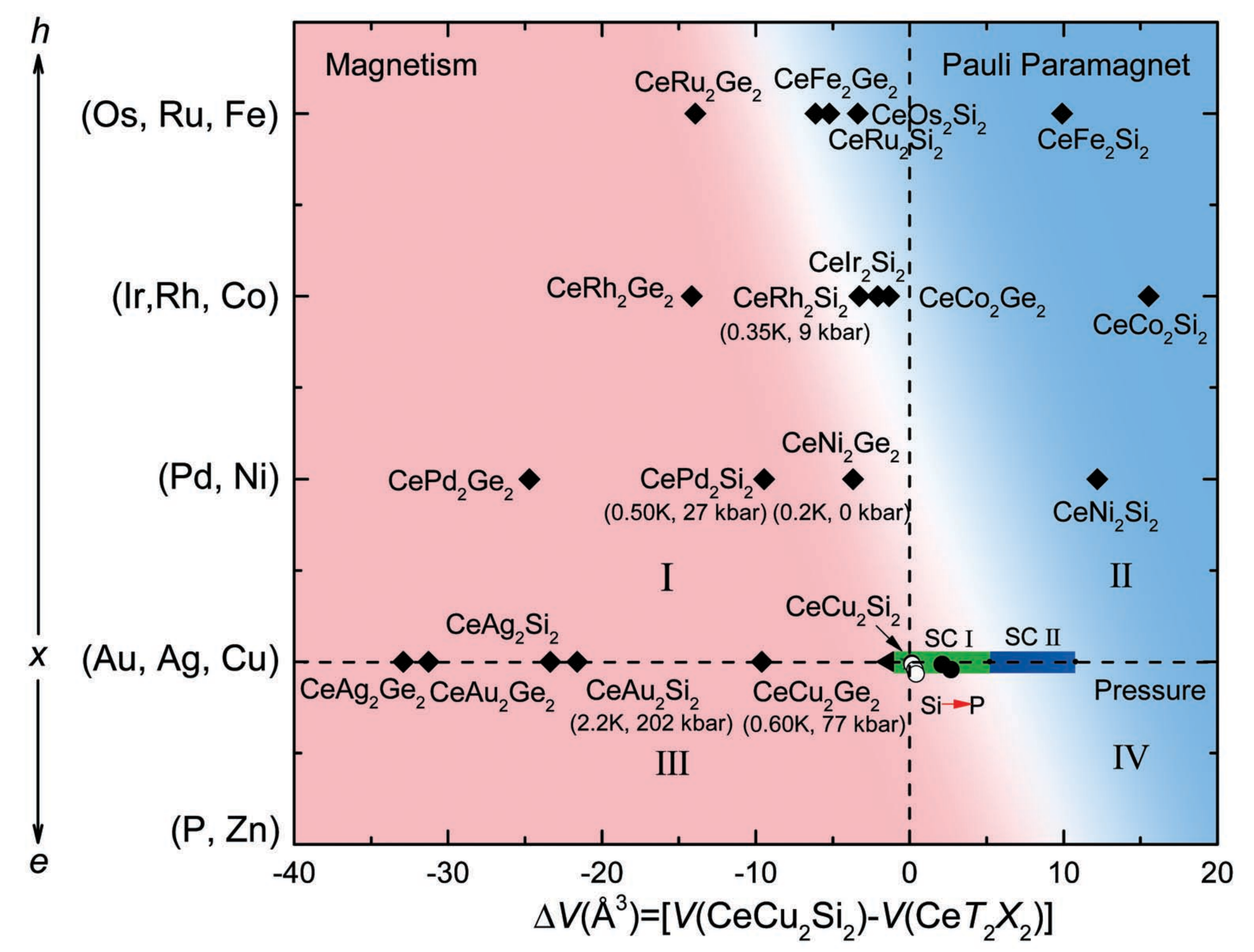}
        \caption{Phase diagram for the compounds Ce$T_2$$X_2$ ($T$ = transition metal and $X$ = Si, Ge) that crystallize in the ThCr$_2$Si$_2$-type structure.~\cite{palstra86,endstra} The axes that define the four quadrants (I-IV) and control the ground state behavior are the difference in unit cell volume ($\Delta$$V$) from that of CeCu$_2$Si$_2$ and increasing $d$-shell filling going from the Fe column to the Cu column ($x$). The white band that traverses the center of the phase diagram approximately separates the magnetic (left hand side) and intermediate valence examples (right hand side). The bars labeled SC1 (green) and SC2 (dark blue) show the regions where superconductivity is observed for CeCu$_2$Si$_2$ along the $\Delta$$V$ axis.~\cite{Steglich79,Knebel96,trovarelli97,yuan03,Yuan06,Steglich12} In parenthesis are the maximum superconducting transition temperatures that are observed under applied pressure for those compounds with an antiferromagnetic quantum phase transition.~\cite{movshovich96,grosche00,grosche01,yuan03,Yuan06,Stockert11,ren14} In quadrant IV the open circles show the zero temperature magnetic phase boundary resulting from the investigation of the substitution series CeCu$_2$(Si$_{1-x}$P$_x$)$_2$ presented here (see Fig.~\ref{fig2}). The closed circles show where applied pressure suppresses magnetic ordering to quantum phase transitions in this series (see Fig.~\ref{fig3}). 
        }
        \label{fig1}
    \end{center}
\end{figure}

We suggest that these insights are broadly relevant to the electronic phase space surrounding CeCu$_2$Si$_2$ (Fig.~\ref{fig1}) where the tuning axes are unit cell volume ($\Delta$$V$) and $s,p,d$-shell filling ($x$). Similar proposals have previously been made,~\cite{delong,koelling,buffat,vildosola} but they mostly do not consider current knowledge about spin/charge instabilities and how they relate to quantum criticality. Thus, it is useful to reexamine the chemical topography with renewed attention paid to the importance of spin and charge fluctuations in critical regions. The isoelectronic subset where $T$ =  Cu, Ag, Au and $X$ =  Si, Ge illustrates the distinct effect of the unit cell volume, where these materials show strengthening of antiferromagnetism with increasing unit cell volume.~\cite{palstra86,endstra} Notably this antiferromagnetism is suppressed with applied pressure, with a return of superconductivity at the magnetic quantum phase transition.~\cite{yuan03,Yuan06,Stockert11,ren14} A similar relationship between magnetism and unit cell volume is seen for the other isoelectronic series ($T$ = Ni, Pd and $X$ = Si, Ge), ($T$ = Co, Rh and $X$ = Si, Ge), and ($T$ = Fe, Ru, Os and $X$ = Si, Ge) [\cite{endstra,palstra86}] and related trends occur in other Ce-based intermetallics, as is expected in the Doniach picture.~\cite{Coleman01,Stewart01,Rosch07,Gegenwart08,Pfleiderer09} The effect of non-isoelectronic tuning is less transparent, but is clarified by considering the examples along the vertical axis with similar unit cell volumes: e.g., those extending from CeCu$_2$Si$_2$ $\rightarrow$ CeNi$_2$Ge$_2$ $\rightarrow$ CeCo$_2$Ge$_2$ $\rightarrow$ CeFe$_2$Ge$_2$ where the $f$-electron valence $v$ tends to increase from $v$ $=$ 3 towards $v$ $\approx$ 3+$\Delta$ between Cu $\rightarrow$ half $d$-shell filling, after which it again decreases towards $v$ $=$ 3.~\cite{delong} This indicates that the hybridization strength is maximized between the Fe and Co columns. A consequence of the relationship between these two distinct axes is that there exists a region that separates the magnetic and nonmagnetic members, where extensive work has shown that clustered along it are those systems that exhibit exotic metallic states, magnetic quantum phase transitions and superconductivity.~\cite{endstra,palstra86,yuan03,Yuan06,movshovich96,grosche00,grosche01,ren14} It is important to note that this picture does not account explicitly for some factors that might sometimes play an important role: e.g., the influences of (1) chemical disorder and (2) differences between 3$d$, 4$d$, and 5$d$ orbitals. Nonetheless, it captures the global trends in this family and our results from Si $\rightarrow$ P substitution indicate that $s/p$ shell filling also fits inside this picture. 

This unifies earlier work and suggests strategies for how to promote behavior such as the high pressure (dome 2) superconductivity in CeCu$_2$Si$_2$ at ambient pressure. For instance, simultaneous doping on both the transition metal and ligand sites, Cu $\rightarrow$ Ni \& Si $\rightarrow$ P, such that their combined effect would be pure unit cell contraction is of interest. However, an important obstacle is chemical disorder, which typically suppresses superconductivity in these materials.~\cite{movshovich96,grosche00,grosche01} An alternative route is to consider electronic tuning in examples that are close to CeCu$_2$Si$_2$ but on the right hand side of the magnetic phase boundary seen in Fig.~\ref{fig1}. For instance, CeNi$_2$Ge$_2$ already exhibits non-Fermi-liquid behavior and incipient superconductivity,~\cite{grosche00} suggesting that small Ge $\rightarrow$ P or As substitution might stabilize bulk superconductivity. We further note that the unstable valence physics and superconductivity extends into quadrant 4, as demonstrated through our applied pressure study. This is the least studied region of the phase diagram and its investigation potentially will yield unexpected discoveries.

\section{Conclusions}
We have studied Si $\rightarrow$ P chemical substitution in CeCu$_2$Si$_2$, which compresses the unit cell volume, weakens the hybridization between the $f$- and conduction electron states, and encourages complex magnetism. At concentrations that show magnetism, applied pressure suppresses the ordering temperature and superconductivity is recovered for samples with low disorder. These results are understood by considering that the electronic hybridization between the $f$- and conduction electrons in this system is controlled by the nearly independent parameters of unit cell volume and $s,p,d$ shell filling. Guided by this topography we have discussed prospects for inducing a valence fluctuation quantum phase transition in the broader family of Ce-based ThCr$_2$Si$_2$-type materials through chemical substitution. 

\section{Acknowledgements}
This work was performed at the National High Magnetic Field Laboratory (NHMFL), which is supported by National Science Foundation Cooperative Agreement No. DMR-1157490, the State of Florida and the DOE." A portion of this work was supported by the  NHMFL User Collaboration Grant Program (UCGP). S. M. Saunders acknowledges support from the NHMFL REU program. We thank Z. Fisk and M. Janoschek for insightful discussions.

% Bibliography
\bibliography{CeCuSirefs}

\begin{thebibliography}{36}%
\makeatletter
\providecommand \@ifxundefined [1]{%
 \@ifx{#1\undefined}
}%
\providecommand \@ifnum [1]{%
 \ifnum #1\expandafter \@firstoftwo
 \else \expandafter \@secondoftwo
 \fi
}%
\providecommand \@ifx [1]{%
 \ifx #1\expandafter \@firstoftwo
 \else \expandafter \@secondoftwo
 \fi
}%
\providecommand \natexlab [1]{#1}%
\providecommand \enquote  [1]{``#1''}%
\providecommand \bibnamefont  [1]{#1}%
\providecommand \bibfnamefont [1]{#1}%
\providecommand \citenamefont [1]{#1}%
\providecommand \href@noop [0]{\@secondoftwo}%
\providecommand \href [0]{\begingroup \@sanitize@url \@href}%
\providecommand \@href[1]{\@@startlink{#1}\@@href}%
\providecommand \@@href[1]{\endgroup#1\@@endlink}%
\providecommand \@sanitize@url [0]{\catcode `\\12\catcode `\$12\catcode
  `\&12\catcode `\#12\catcode `\^12\catcode `\_12\catcode `\%12\relax}%
\providecommand \@@startlink[1]{}%
\providecommand \@@endlink[0]{}%
\providecommand \url  [0]{\begingroup\@sanitize@url \@url }%
\providecommand \@url [1]{\endgroup\@href {#1}{\urlprefix }}%
\providecommand \urlprefix  [0]{URL }%
\providecommand \Eprint [0]{\href }%
\providecommand \doibase [0]{http://dx.doi.org/}%
\providecommand \selectlanguage [0]{\@gobble}%
\providecommand \bibinfo  [0]{\@secondoftwo}%
\providecommand \bibfield  [0]{\@secondoftwo}%
\providecommand \translation [1]{[#1]}%
\providecommand \BibitemOpen [0]{}%
\providecommand \bibitemStop [0]{}%
\providecommand \bibitemNoStop [0]{.\EOS\space}%
\providecommand \EOS [0]{\spacefactor3000\relax}%
\providecommand \BibitemShut  [1]{\csname bibitem#1\endcsname}%
\let\auto@bib@innerbib\@empty
%</preamble>
\bibitem [{\citenamefont {Mott}(1952)}]{mott}%
  \BibitemOpen
  \bibfield  {author} {\bibinfo {author} {\bibfnamefont {N.}~\bibnamefont
  {Mott}},\ }\href@noop {} {\emph {\bibinfo {title} {Progress in Metal Physics
  3}}}\ (\bibinfo  {publisher} {London Pergamon Press},\ \bibinfo {year}
  {1952})\BibitemShut {NoStop}%
\bibitem [{\citenamefont {Herring}(1960)}]{herring}%
  \BibitemOpen
  \bibfield  {author} {\bibinfo {author} {\bibfnamefont {C.}~\bibnamefont
  {Herring}},\ }\href@noop {} {\bibfield  {journal} {\bibinfo  {journal}
  {Journal of Applied Physics}\ }\textbf {\bibinfo {volume} {31}},\ \bibinfo
  {pages} {S3} (\bibinfo {year} {1960})}\BibitemShut {NoStop}%
\bibitem [{\citenamefont {Coleman}\ \emph {et~al.}(2001)\citenamefont
  {Coleman}, \citenamefont {P{\'e}pin}, \citenamefont {Si},\ and\ \citenamefont
  {Ramazashvili}}]{Coleman01}%
  \BibitemOpen
  \bibfield  {author} {\bibinfo {author} {\bibfnamefont {P.}~\bibnamefont
  {Coleman}}, \bibinfo {author} {\bibfnamefont {C.}~\bibnamefont {P{\'e}pin}},
  \bibinfo {author} {\bibfnamefont {Q.}~\bibnamefont {Si}}, \ and\ \bibinfo
  {author} {\bibfnamefont {R.}~\bibnamefont {Ramazashvili}},\ }\href@noop {}
  {\bibfield  {journal} {\bibinfo  {journal} {Journal of Physics: Condensed
  Matter}\ }\textbf {\bibinfo {volume} {13}},\ \bibinfo {pages} {R723}
  (\bibinfo {year} {2001})}\BibitemShut {NoStop}%
\bibitem [{\citenamefont {Stewart}(2001)}]{Stewart01}%
  \BibitemOpen
  \bibfield  {author} {\bibinfo {author} {\bibfnamefont {G.~R.}\ \bibnamefont
  {Stewart}},\ }\href@noop {} {\bibfield  {journal} {\bibinfo  {journal}
  {Reviews of Modern Physics}\ }\textbf {\bibinfo {volume} {73}},\ \bibinfo
  {pages} {797} (\bibinfo {year} {2001})}\BibitemShut {NoStop}%
\bibitem [{\citenamefont {v.~L{\"o}hneysen}\ \emph {et~al.}(2007)\citenamefont
  {v.~L{\"o}hneysen}, \citenamefont {Rosch}, \citenamefont {Vojta},\ and\
  \citenamefont {W{\"o}lfle}}]{Rosch07}%
  \BibitemOpen
  \bibfield  {author} {\bibinfo {author} {\bibfnamefont {H.}~\bibnamefont
  {v.~L{\"o}hneysen}}, \bibinfo {author} {\bibfnamefont {A.}~\bibnamefont
  {Rosch}}, \bibinfo {author} {\bibfnamefont {M.}~\bibnamefont {Vojta}}, \ and\
  \bibinfo {author} {\bibfnamefont {P.}~\bibnamefont {W{\"o}lfle}},\
  }\href@noop {} {\bibfield  {journal} {\bibinfo  {journal} {Reviews of Modern
  Physics}\ }\textbf {\bibinfo {volume} {79}},\ \bibinfo {pages} {1015}
  (\bibinfo {year} {2007})}\BibitemShut {NoStop}%
\bibitem [{\citenamefont {Gegenwart}\ \emph {et~al.}(2008)\citenamefont
  {Gegenwart}, \citenamefont {Si},\ and\ \citenamefont
  {Steglich}}]{Gegenwart08}%
  \BibitemOpen
  \bibfield  {author} {\bibinfo {author} {\bibfnamefont {P.}~\bibnamefont
  {Gegenwart}}, \bibinfo {author} {\bibfnamefont {Q.}~\bibnamefont {Si}}, \
  and\ \bibinfo {author} {\bibfnamefont {F.}~\bibnamefont {Steglich}},\
  }\href@noop {} {\bibfield  {journal} {\bibinfo  {journal} {Nature Physics}\
  }\textbf {\bibinfo {volume} {4}},\ \bibinfo {pages} {186} (\bibinfo {year}
  {2008})}\BibitemShut {NoStop}%
\bibitem [{\citenamefont {Pfleiderer}(2009)}]{Pfleiderer09}%
  \BibitemOpen
  \bibfield  {author} {\bibinfo {author} {\bibfnamefont {C.}~\bibnamefont
  {Pfleiderer}},\ }\href@noop {} {\bibfield  {journal} {\bibinfo  {journal}
  {Reviews of Modern Physics}\ }\textbf {\bibinfo {volume} {81}},\ \bibinfo
  {pages} {1551} (\bibinfo {year} {2009})}\BibitemShut {NoStop}%
\bibitem [{\citenamefont {Doniach}(1977)}]{Doniach_77}%
  \BibitemOpen
  \bibfield  {author} {\bibinfo {author} {\bibfnamefont {S.}~\bibnamefont
  {Doniach}},\ }\href@noop {} {\bibfield  {journal} {\bibinfo  {journal}
  {Physica B + C}\ }\textbf {\bibinfo {volume} {91}},\ \bibinfo {pages} {231}
  (\bibinfo {year} {1977})}\BibitemShut {NoStop}%
\bibitem [{\citenamefont {Kondo}(1964)}]{kondo}%
  \BibitemOpen
  \bibfield  {author} {\bibinfo {author} {\bibfnamefont {J.}~\bibnamefont
  {Kondo}},\ }\href@noop {} {\bibfield  {journal} {\bibinfo  {journal}
  {Progress in Theoretical Physics}\ }\textbf {\bibinfo {volume} {32}},\
  \bibinfo {pages} {37} (\bibinfo {year} {1964})}\BibitemShut {NoStop}%
\bibitem [{\citenamefont {Ruderman}\ and\ \citenamefont
  {Kittel}(1954)}]{ruderman}%
  \BibitemOpen
  \bibfield  {author} {\bibinfo {author} {\bibfnamefont {M.~A.}\ \bibnamefont
  {Ruderman}}\ and\ \bibinfo {author} {\bibfnamefont {C.}~\bibnamefont
  {Kittel}},\ }\href@noop {} {\bibfield  {journal} {\bibinfo  {journal}
  {Physical Review}\ }\textbf {\bibinfo {volume} {96}},\ \bibinfo {pages} {99}
  (\bibinfo {year} {1954})}\BibitemShut {NoStop}%
\bibitem [{\citenamefont {Kasuya}(1956)}]{kasuya}%
  \BibitemOpen
  \bibfield  {author} {\bibinfo {author} {\bibfnamefont {T.}~\bibnamefont
  {Kasuya}},\ }\href@noop {} {\bibfield  {journal} {\bibinfo  {journal}
  {Progress in Theoretical Physics}\ }\textbf {\bibinfo {volume} {16}},\
  \bibinfo {pages} {45} (\bibinfo {year} {1956})}\BibitemShut {NoStop}%
\bibitem [{\citenamefont {Yosida}(1957)}]{yosida}%
  \BibitemOpen
  \bibfield  {author} {\bibinfo {author} {\bibfnamefont {K.}~\bibnamefont
  {Yosida}},\ }\href@noop {} {\bibfield  {journal} {\bibinfo  {journal}
  {Physical Review}\ }\textbf {\bibinfo {volume} {106}},\ \bibinfo {pages}
  {893} (\bibinfo {year} {1957})}\BibitemShut {NoStop}%
\bibitem [{\citenamefont {Thompson}\ and\ \citenamefont
  {Fisk}(2012)}]{thompson12}%
  \BibitemOpen
  \bibfield  {author} {\bibinfo {author} {\bibfnamefont {J.~D.}\ \bibnamefont
  {Thompson}}\ and\ \bibinfo {author} {\bibfnamefont {Z.}~\bibnamefont
  {Fisk}},\ }\href@noop {} {\bibfield  {journal} {\bibinfo  {journal} {Journal
  of the Physical Society of Japan}\ }\textbf {\bibinfo {volume} {81}},\
  \bibinfo {pages} {011002} (\bibinfo {year} {2012})}\BibitemShut {NoStop}%
\bibitem [{\citenamefont {Gofryk}\ \emph {et~al.}(2012)\citenamefont {Gofryk},
  \citenamefont {Ronning}, \citenamefont {Zhu}, \citenamefont {Ou},
  \citenamefont {Tobash}, \citenamefont {Stoyko}, \citenamefont {Lu},
  \citenamefont {Mar}, \citenamefont {Park}, \citenamefont {Bauer},
  \citenamefont {Thompson},\ and\ \citenamefont {Fisk}}]{gofryk12}%
  \BibitemOpen
  \bibfield  {author} {\bibinfo {author} {\bibfnamefont {K.}~\bibnamefont
  {Gofryk}}, \bibinfo {author} {\bibfnamefont {F.}~\bibnamefont {Ronning}},
  \bibinfo {author} {\bibfnamefont {J.-X.}\ \bibnamefont {Zhu}}, \bibinfo
  {author} {\bibfnamefont {M.~N.}\ \bibnamefont {Ou}}, \bibinfo {author}
  {\bibfnamefont {P.~H.}\ \bibnamefont {Tobash}}, \bibinfo {author}
  {\bibfnamefont {S.~S.}\ \bibnamefont {Stoyko}}, \bibinfo {author}
  {\bibfnamefont {X.}~\bibnamefont {Lu}}, \bibinfo {author} {\bibfnamefont
  {A.}~\bibnamefont {Mar}}, \bibinfo {author} {\bibfnamefont {T.}~\bibnamefont
  {Park}}, \bibinfo {author} {\bibfnamefont {E.~D.}\ \bibnamefont {Bauer}},
  \bibinfo {author} {\bibfnamefont {J.~D.}\ \bibnamefont {Thompson}}, \ and\
  \bibinfo {author} {\bibfnamefont {Z.}~\bibnamefont {Fisk}},\ }\href@noop {}
  {\bibfield  {journal} {\bibinfo  {journal} {Physical Review Letters}\
  }\textbf {\bibinfo {volume} {109}},\ \bibinfo {pages} {186402} (\bibinfo
  {year} {2012})}\BibitemShut {NoStop}%
\bibitem [{\citenamefont {Palstra}\ \emph {et~al.}(1986)\citenamefont
  {Palstra}, \citenamefont {Menovsky}, \citenamefont {Nieuwenhuys},\ and\
  \citenamefont {Mydosh}}]{palstra86}%
  \BibitemOpen
  \bibfield  {author} {\bibinfo {author} {\bibfnamefont {T.~T.~M.}\
  \bibnamefont {Palstra}}, \bibinfo {author} {\bibfnamefont {A.~A.}\
  \bibnamefont {Menovsky}}, \bibinfo {author} {\bibfnamefont {G.~J.}\
  \bibnamefont {Nieuwenhuys}}, \ and\ \bibinfo {author} {\bibfnamefont {J.~A.}\
  \bibnamefont {Mydosh}},\ }\href@noop {} {\bibfield  {journal} {\bibinfo
  {journal} {Journal of Magnetism and Magnetic Materials}\ }\textbf {\bibinfo
  {volume} {54-57(Part1)}},\ \bibinfo {pages} {435} (\bibinfo {year}
  {1986})}\BibitemShut {NoStop}%
\bibitem [{\citenamefont {Endstra}\ \emph {et~al.}(1993)\citenamefont
  {Endstra}, \citenamefont {Nieuwenhuys}, \citenamefont {Palstra},\ and\
  \citenamefont {Mydosh}}]{endstra}%
  \BibitemOpen
  \bibfield  {author} {\bibinfo {author} {\bibfnamefont {T.}~\bibnamefont
  {Endstra}}, \bibinfo {author} {\bibfnamefont {G.~J.}\ \bibnamefont
  {Nieuwenhuys}}, \bibinfo {author} {\bibfnamefont {T.~T.~M.}\ \bibnamefont
  {Palstra}}, \ and\ \bibinfo {author} {\bibfnamefont {J.~A.}\ \bibnamefont
  {Mydosh}},\ }\href@noop {} {\bibfield  {journal} {\bibinfo  {journal}
  {Physical Review B}\ }\textbf {\bibinfo {volume} {48}},\ \bibinfo {pages}
  {9595} (\bibinfo {year} {1993})}\BibitemShut {NoStop}%
\bibitem [{\citenamefont {Steglich}\ \emph {et~al.}(1979)\citenamefont
  {Steglich}, \citenamefont {Aarts}, \citenamefont {Bredl}, \citenamefont
  {Lieke}, \citenamefont {Meschede}, \citenamefont {Franz},\ and\ \citenamefont
  {Sch\"afer}}]{Steglich79}%
  \BibitemOpen
  \bibfield  {author} {\bibinfo {author} {\bibfnamefont {F.}~\bibnamefont
  {Steglich}}, \bibinfo {author} {\bibfnamefont {J.}~\bibnamefont {Aarts}},
  \bibinfo {author} {\bibfnamefont {C.~D.}\ \bibnamefont {Bredl}}, \bibinfo
  {author} {\bibfnamefont {W.}~\bibnamefont {Lieke}}, \bibinfo {author}
  {\bibfnamefont {D.}~\bibnamefont {Meschede}}, \bibinfo {author}
  {\bibfnamefont {W.}~\bibnamefont {Franz}}, \ and\ \bibinfo {author}
  {\bibfnamefont {H.}~\bibnamefont {Sch\"afer}},\ }\href@noop {} {\bibfield
  {journal} {\bibinfo  {journal} {Physical Review Letters}\ }\textbf {\bibinfo
  {volume} {43}},\ \bibinfo {pages} {1892} (\bibinfo {year}
  {1979})}\BibitemShut {NoStop}%
\bibitem [{\citenamefont {Knebel}\ \emph {et~al.}(1996)\citenamefont {Knebel},
  \citenamefont {Eggert}, \citenamefont {Engelmann}, \citenamefont {Viana},
  \citenamefont {Krimmel}, \citenamefont {Dressel},\ and\ \citenamefont
  {Loidl}}]{Knebel96}%
  \BibitemOpen
  \bibfield  {author} {\bibinfo {author} {\bibfnamefont {G.}~\bibnamefont
  {Knebel}}, \bibinfo {author} {\bibfnamefont {C.}~\bibnamefont {Eggert}},
  \bibinfo {author} {\bibfnamefont {D.}~\bibnamefont {Engelmann}}, \bibinfo
  {author} {\bibfnamefont {R.}~\bibnamefont {Viana}}, \bibinfo {author}
  {\bibfnamefont {A.}~\bibnamefont {Krimmel}}, \bibinfo {author} {\bibfnamefont
  {M.}~\bibnamefont {Dressel}}, \ and\ \bibinfo {author} {\bibfnamefont
  {A.}~\bibnamefont {Loidl}},\ }\href@noop {} {\bibfield  {journal} {\bibinfo
  {journal} {Physical Review B}\ }\textbf {\bibinfo {volume} {53}},\ \bibinfo
  {pages} {11586} (\bibinfo {year} {1996})}\BibitemShut {NoStop}%
\bibitem [{\citenamefont {Trovarelli}\ \emph {et~al.}(1997)\citenamefont
  {Trovarelli}, \citenamefont {Weiden}, \citenamefont {M{\"u}ller-Reisner},
  \citenamefont {G\'omez-Berisso}, \citenamefont {Gegenwart}, \citenamefont
  {Deppe}, \citenamefont {Geibel}, \citenamefont {Sereni},\ and\ \citenamefont
  {Steglich}}]{trovarelli97}%
  \BibitemOpen
  \bibfield  {author} {\bibinfo {author} {\bibfnamefont {O.}~\bibnamefont
  {Trovarelli}}, \bibinfo {author} {\bibfnamefont {M.}~\bibnamefont {Weiden}},
  \bibinfo {author} {\bibfnamefont {R.}~\bibnamefont {M{\"u}ller-Reisner}},
  \bibinfo {author} {\bibfnamefont {M.}~\bibnamefont {G\'omez-Berisso}},
  \bibinfo {author} {\bibfnamefont {P.}~\bibnamefont {Gegenwart}}, \bibinfo
  {author} {\bibfnamefont {M.}~\bibnamefont {Deppe}}, \bibinfo {author}
  {\bibfnamefont {C.}~\bibnamefont {Geibel}}, \bibinfo {author} {\bibfnamefont
  {J.~G.}\ \bibnamefont {Sereni}}, \ and\ \bibinfo {author} {\bibfnamefont
  {F.}~\bibnamefont {Steglich}},\ }\href@noop {} {\bibfield  {journal}
  {\bibinfo  {journal} {Physical Review B}\ }\textbf {\bibinfo {volume} {56}},\
  \bibinfo {pages} {678} (\bibinfo {year} {1997})}\BibitemShut {NoStop}%
\bibitem [{\citenamefont {Yuan}\ \emph {et~al.}(2003)\citenamefont {Yuan},
  \citenamefont {Grosche}, \citenamefont {Deppe}, \citenamefont {Geibel},
  \citenamefont {Sparn},\ and\ \citenamefont {Steglich}}]{yuan03}%
  \BibitemOpen
  \bibfield  {author} {\bibinfo {author} {\bibfnamefont {H.~Q.}\ \bibnamefont
  {Yuan}}, \bibinfo {author} {\bibfnamefont {F.~M.}\ \bibnamefont {Grosche}},
  \bibinfo {author} {\bibfnamefont {M.}~\bibnamefont {Deppe}}, \bibinfo
  {author} {\bibfnamefont {C.}~\bibnamefont {Geibel}}, \bibinfo {author}
  {\bibfnamefont {G.}~\bibnamefont {Sparn}}, \ and\ \bibinfo {author}
  {\bibfnamefont {F.}~\bibnamefont {Steglich}},\ }\href@noop {} {\bibfield
  {journal} {\bibinfo  {journal} {Science}\ }\textbf {\bibinfo {volume}
  {302}},\ \bibinfo {pages} {2104} (\bibinfo {year} {2003})}\BibitemShut
  {NoStop}%
\bibitem [{\citenamefont {Yuan}\ \emph {et~al.}(2006)\citenamefont {Yuan},
  \citenamefont {Grosche}, \citenamefont {Deppe}, \citenamefont {Sparn},
  \citenamefont {Geibel},\ and\ \citenamefont {Steglich}}]{Yuan06}%
  \BibitemOpen
  \bibfield  {author} {\bibinfo {author} {\bibfnamefont {H.~Q.}\ \bibnamefont
  {Yuan}}, \bibinfo {author} {\bibfnamefont {F.~M.}\ \bibnamefont {Grosche}},
  \bibinfo {author} {\bibfnamefont {M.}~\bibnamefont {Deppe}}, \bibinfo
  {author} {\bibfnamefont {G.}~\bibnamefont {Sparn}}, \bibinfo {author}
  {\bibfnamefont {C.}~\bibnamefont {Geibel}}, \ and\ \bibinfo {author}
  {\bibfnamefont {F.}~\bibnamefont {Steglich}},\ }\href@noop {} {\bibfield
  {journal} {\bibinfo  {journal} {Physical Review Letters}\ }\textbf {\bibinfo
  {volume} {96}},\ \bibinfo {pages} {047008} (\bibinfo {year}
  {2006})}\BibitemShut {NoStop}%
\bibitem [{\citenamefont {Holmes}\ \emph {et~al.}(2007)\citenamefont {Holmes},
  \citenamefont {Jaccard},\ and\ \citenamefont {Miyake}}]{holmes07}%
  \BibitemOpen
  \bibfield  {author} {\bibinfo {author} {\bibfnamefont {A.~T.}\ \bibnamefont
  {Holmes}}, \bibinfo {author} {\bibfnamefont {D.}~\bibnamefont {Jaccard}}, \
  and\ \bibinfo {author} {\bibfnamefont {M.}~\bibnamefont {Miyake}},\
  }\href@noop {} {\bibfield  {journal} {\bibinfo  {journal} {Journal of the
  Physical Society of Japan}\ }\textbf {\bibinfo {volume} {76}},\ \bibinfo
  {pages} {051002} (\bibinfo {year} {2007})}\BibitemShut {NoStop}%
\bibitem [{\citenamefont {Steglich}(2012)}]{Steglich12}%
  \BibitemOpen
  \bibfield  {author} {\bibinfo {author} {\bibfnamefont {F.}~\bibnamefont
  {Steglich}},\ }\href@noop {} {\bibfield  {journal} {\bibinfo  {journal}
  {Journal of Physics: Conference Series}\ }\textbf {\bibinfo {volume} {400}},\
  \bibinfo {pages} {022111} (\bibinfo {year} {2012})}\BibitemShut {NoStop}%
\bibitem [{\citenamefont {Kobayashi}\ \emph {et~al.}(2013)\citenamefont
  {Kobayashi}, \citenamefont {Fujiwara}, \citenamefont {Takeda}, \citenamefont
  {Harima}, \citenamefont {Ikeda}, \citenamefont {Adachi}, \citenamefont
  {Ohishi}, \citenamefont {Geibel},\ and\ \citenamefont
  {Steglich}}]{kobayashi13}%
  \BibitemOpen
  \bibfield  {author} {\bibinfo {author} {\bibfnamefont {T.~C.}\ \bibnamefont
  {Kobayashi}}, \bibinfo {author} {\bibfnamefont {K.}~\bibnamefont {Fujiwara}},
  \bibinfo {author} {\bibfnamefont {K.}~\bibnamefont {Takeda}}, \bibinfo
  {author} {\bibfnamefont {H.}~\bibnamefont {Harima}}, \bibinfo {author}
  {\bibfnamefont {Y.}~\bibnamefont {Ikeda}}, \bibinfo {author} {\bibfnamefont
  {T.}~\bibnamefont {Adachi}}, \bibinfo {author} {\bibfnamefont
  {Y.}~\bibnamefont {Ohishi}}, \bibinfo {author} {\bibfnamefont
  {C.}~\bibnamefont {Geibel}}, \ and\ \bibinfo {author} {\bibfnamefont
  {F.}~\bibnamefont {Steglich}},\ }\href@noop {} {\bibfield  {journal}
  {\bibinfo  {journal} {Journal of the Physical Society of Japan}\ }\textbf
  {\bibinfo {volume} {82}},\ \bibinfo {pages} {114701} (\bibinfo {year}
  {2013})}\BibitemShut {NoStop}%
\bibitem [{\citenamefont {Rueff}\ \emph {et~al.}(2011)\citenamefont {Rueff},
  \citenamefont {Raymond}, \citenamefont {Taguchi}, \citenamefont {Sikora},
  \citenamefont {Iti{\'e}}, \citenamefont {Baudelet}, \citenamefont
  {Braithwaite}, \citenamefont {Knebel},\ and\ \citenamefont
  {Jaccard}}]{rueff}%
  \BibitemOpen
  \bibfield  {author} {\bibinfo {author} {\bibfnamefont {J.-P.}\ \bibnamefont
  {Rueff}}, \bibinfo {author} {\bibfnamefont {S.}~\bibnamefont {Raymond}},
  \bibinfo {author} {\bibfnamefont {M.}~\bibnamefont {Taguchi}}, \bibinfo
  {author} {\bibfnamefont {M.}~\bibnamefont {Sikora}}, \bibinfo {author}
  {\bibfnamefont {J.-P.}\ \bibnamefont {Iti{\'e}}}, \bibinfo {author}
  {\bibfnamefont {F.}~\bibnamefont {Baudelet}}, \bibinfo {author}
  {\bibfnamefont {D.}~\bibnamefont {Braithwaite}}, \bibinfo {author}
  {\bibfnamefont {G.}~\bibnamefont {Knebel}}, \ and\ \bibinfo {author}
  {\bibfnamefont {D.}~\bibnamefont {Jaccard}},\ }\href@noop {} {\bibfield
  {journal} {\bibinfo  {journal} {Physical Review Letters}\ }\textbf {\bibinfo
  {volume} {106}},\ \bibinfo {pages} {186405} (\bibinfo {year}
  {2011})}\BibitemShut {NoStop}%
\bibitem [{\citenamefont {Miyake}(2007)}]{miyake_07}%
  \BibitemOpen
  \bibfield  {author} {\bibinfo {author} {\bibfnamefont {K.}~\bibnamefont
  {Miyake}},\ }\href@noop {} {\bibfield  {journal} {\bibinfo  {journal}
  {Journal of Physics: Condensed Matter}\ }\textbf {\bibinfo {volume} {19}},\
  \bibinfo {pages} {125201} (\bibinfo {year} {2007})}\BibitemShut {NoStop}%
\bibitem [{\citenamefont {Spain}\ \emph {et~al.}(1986)\citenamefont {Spain},
  \citenamefont {Rauchschwalbe},\ and\ \citenamefont {Hochheimer}}]{spain}%
  \BibitemOpen
  \bibfield  {author} {\bibinfo {author} {\bibfnamefont {I.~L.}\ \bibnamefont
  {Spain}}, \bibinfo {author} {\bibfnamefont {F.~S.~S.}\ \bibnamefont
  {Rauchschwalbe}}, \ and\ \bibinfo {author} {\bibfnamefont {H.~D.}\
  \bibnamefont {Hochheimer}},\ }\href@noop {} {\bibfield  {journal} {\bibinfo
  {journal} {Physica B + C}\ }\textbf {\bibinfo {volume} {139-140}},\ \bibinfo
  {pages} {449} (\bibinfo {year} {1986})}\BibitemShut {NoStop}%
\bibitem [{\citenamefont {Movshovich}\ \emph {et~al.}(1996)\citenamefont
  {Movshovich}, \citenamefont {Graf}, \citenamefont {Mandrus}, \citenamefont
  {Thompson}, \citenamefont {Smith},\ and\ \citenamefont
  {Fisk}}]{movshovich96}%
  \BibitemOpen
  \bibfield  {author} {\bibinfo {author} {\bibfnamefont {R.}~\bibnamefont
  {Movshovich}}, \bibinfo {author} {\bibfnamefont {T.}~\bibnamefont {Graf}},
  \bibinfo {author} {\bibfnamefont {D.}~\bibnamefont {Mandrus}}, \bibinfo
  {author} {\bibfnamefont {J.~D.}\ \bibnamefont {Thompson}}, \bibinfo {author}
  {\bibfnamefont {J.~L.}\ \bibnamefont {Smith}}, \ and\ \bibinfo {author}
  {\bibfnamefont {Z.}~\bibnamefont {Fisk}},\ }\href@noop {} {\bibfield
  {journal} {\bibinfo  {journal} {Physical Review B}\ }\textbf {\bibinfo
  {volume} {53}},\ \bibinfo {pages} {8241} (\bibinfo {year}
  {1996})}\BibitemShut {NoStop}%
\bibitem [{\citenamefont {Grosche}\ \emph {et~al.}(2000)\citenamefont
  {Grosche}, \citenamefont {Agarwal}, \citenamefont {Julian}, \citenamefont
  {Wilson}, \citenamefont {Haselwimmer}, \citenamefont {Lister}, \citenamefont
  {Mathur}, \citenamefont {Carter}, \citenamefont {Saxena},\ and\ \citenamefont
  {Lonzarich}}]{grosche00}%
  \BibitemOpen
  \bibfield  {author} {\bibinfo {author} {\bibfnamefont {F.~M.}\ \bibnamefont
  {Grosche}}, \bibinfo {author} {\bibfnamefont {P.}~\bibnamefont {Agarwal}},
  \bibinfo {author} {\bibfnamefont {S.~R.}\ \bibnamefont {Julian}}, \bibinfo
  {author} {\bibfnamefont {N.~J.}\ \bibnamefont {Wilson}}, \bibinfo {author}
  {\bibfnamefont {R.~K.~W.}\ \bibnamefont {Haselwimmer}}, \bibinfo {author}
  {\bibfnamefont {S.~J.~S.}\ \bibnamefont {Lister}}, \bibinfo {author}
  {\bibfnamefont {N.~D.}\ \bibnamefont {Mathur}}, \bibinfo {author}
  {\bibfnamefont {F.~V.}\ \bibnamefont {Carter}}, \bibinfo {author}
  {\bibfnamefont {S.~S.}\ \bibnamefont {Saxena}}, \ and\ \bibinfo {author}
  {\bibfnamefont {G.~G.}\ \bibnamefont {Lonzarich}},\ }\href@noop {} {\bibfield
   {journal} {\bibinfo  {journal} {Journal of Physics: Condensed Matter}\
  }\textbf {\bibinfo {volume} {12}},\ \bibinfo {pages} {32} (\bibinfo {year}
  {2000})}\BibitemShut {NoStop}%
\bibitem [{\citenamefont {Grosche}\ \emph {et~al.}(2001)\citenamefont
  {Grosche}, \citenamefont {Walker}, \citenamefont {Julian}, \citenamefont
  {Mathur}, \citenamefont {Freye}, \citenamefont {Steiner},\ and\ \citenamefont
  {Lonzarich}}]{grosche01}%
  \BibitemOpen
  \bibfield  {author} {\bibinfo {author} {\bibfnamefont {F.~M.}\ \bibnamefont
  {Grosche}}, \bibinfo {author} {\bibfnamefont {I.~R.}\ \bibnamefont {Walker}},
  \bibinfo {author} {\bibfnamefont {S.~R.}\ \bibnamefont {Julian}}, \bibinfo
  {author} {\bibfnamefont {N.~D.}\ \bibnamefont {Mathur}}, \bibinfo {author}
  {\bibfnamefont {D.~M.}\ \bibnamefont {Freye}}, \bibinfo {author}
  {\bibfnamefont {M.~J.}\ \bibnamefont {Steiner}}, \ and\ \bibinfo {author}
  {\bibfnamefont {G.~G.}\ \bibnamefont {Lonzarich}},\ }\href@noop {} {\bibfield
   {journal} {\bibinfo  {journal} {Journal of Physics: Condensed Matter}\
  }\textbf {\bibinfo {volume} {13}},\ \bibinfo {pages} {2845} (\bibinfo {year}
  {2001})}\BibitemShut {NoStop}%
\bibitem [{\citenamefont {Stockert}\ \emph {et~al.}(2011)\citenamefont
  {Stockert}, \citenamefont {Arndt}, \citenamefont {Faulhaber}, \citenamefont
  {Geibel}, \citenamefont {Jeevan}, \citenamefont {Kirchner}, \citenamefont
  {Loewenhaupt}, \citenamefont {Schmalzl}, \citenamefont {Schmidt},
  \citenamefont {Si},\ and\ \citenamefont {Steglich}}]{Stockert11}%
  \BibitemOpen
  \bibfield  {author} {\bibinfo {author} {\bibfnamefont {O.}~\bibnamefont
  {Stockert}}, \bibinfo {author} {\bibfnamefont {J.}~\bibnamefont {Arndt}},
  \bibinfo {author} {\bibfnamefont {E.}~\bibnamefont {Faulhaber}}, \bibinfo
  {author} {\bibfnamefont {C.}~\bibnamefont {Geibel}}, \bibinfo {author}
  {\bibfnamefont {H.~S.}\ \bibnamefont {Jeevan}}, \bibinfo {author}
  {\bibfnamefont {S.}~\bibnamefont {Kirchner}}, \bibinfo {author}
  {\bibfnamefont {M.}~\bibnamefont {Loewenhaupt}}, \bibinfo {author}
  {\bibfnamefont {K.}~\bibnamefont {Schmalzl}}, \bibinfo {author}
  {\bibfnamefont {W.}~\bibnamefont {Schmidt}}, \bibinfo {author} {\bibfnamefont
  {Q.}~\bibnamefont {Si}}, \ and\ \bibinfo {author} {\bibfnamefont
  {F.}~\bibnamefont {Steglich}},\ }\href@noop {} {\bibfield  {journal}
  {\bibinfo  {journal} {Nature Physics}\ }\textbf {\bibinfo {volume} {7}},\
  \bibinfo {pages} {119} (\bibinfo {year} {2011})}\BibitemShut {NoStop}%
\bibitem [{\citenamefont {Ren}\ \emph {et~al.}(2014)\citenamefont {Ren},
  \citenamefont {Pourovskii}, \citenamefont {Giriat}, \citenamefont {Lapertot},
  \citenamefont {Georges},\ and\ \citenamefont {Jaccard}}]{ren14}%
  \BibitemOpen
  \bibfield  {author} {\bibinfo {author} {\bibfnamefont {Z.}~\bibnamefont
  {Ren}}, \bibinfo {author} {\bibfnamefont {L.~V.}\ \bibnamefont {Pourovskii}},
  \bibinfo {author} {\bibfnamefont {G.}~\bibnamefont {Giriat}}, \bibinfo
  {author} {\bibfnamefont {G.}~\bibnamefont {Lapertot}}, \bibinfo {author}
  {\bibfnamefont {A.}~\bibnamefont {Georges}}, \ and\ \bibinfo {author}
  {\bibfnamefont {D.}~\bibnamefont {Jaccard}},\ }\href@noop {} {\bibfield
  {journal} {\bibinfo  {journal} {Physical Review X}\ }\textbf {\bibinfo
  {volume} {4}},\ \bibinfo {pages} {031055} (\bibinfo {year}
  {2014})}\BibitemShut {NoStop}%
\bibitem [{\citenamefont {Niefeld}\ \emph {et~al.}(1985)\citenamefont
  {Niefeld}, \citenamefont {Croft}, \citenamefont {Mihalisin}, \citenamefont
  {Segre}, \citenamefont {Madigan}, \citenamefont {Torikachvili}, \citenamefont
  {Maple},\ and\ \citenamefont {DeLong}}]{delong}%
  \BibitemOpen
  \bibfield  {author} {\bibinfo {author} {\bibfnamefont {R.~A.}\ \bibnamefont
  {Niefeld}}, \bibinfo {author} {\bibfnamefont {M.}~\bibnamefont {Croft}},
  \bibinfo {author} {\bibfnamefont {T.}~\bibnamefont {Mihalisin}}, \bibinfo
  {author} {\bibfnamefont {C.~U.}\ \bibnamefont {Segre}}, \bibinfo {author}
  {\bibfnamefont {M.}~\bibnamefont {Madigan}}, \bibinfo {author} {\bibfnamefont
  {M.~S.}\ \bibnamefont {Torikachvili}}, \bibinfo {author} {\bibfnamefont
  {M.~B.}\ \bibnamefont {Maple}}, \ and\ \bibinfo {author} {\bibfnamefont
  {L.~E.}\ \bibnamefont {DeLong}},\ }\href@noop {} {\bibfield  {journal}
  {\bibinfo  {journal} {Physical Review B}\ }\textbf {\bibinfo {volume} {32}},\
  \bibinfo {pages} {6928} (\bibinfo {year} {1985})}\BibitemShut {NoStop}%
\bibitem [{\citenamefont {Koelling}\ \emph {et~al.}(1985)\citenamefont
  {Koelling}, \citenamefont {Dunlap},\ and\ \citenamefont
  {Crabtree}}]{koelling}%
  \BibitemOpen
  \bibfield  {author} {\bibinfo {author} {\bibfnamefont {D.~D.}\ \bibnamefont
  {Koelling}}, \bibinfo {author} {\bibfnamefont {B.~D.}\ \bibnamefont
  {Dunlap}}, \ and\ \bibinfo {author} {\bibfnamefont {G.~W.}\ \bibnamefont
  {Crabtree}},\ }\href@noop {} {\bibfield  {journal} {\bibinfo  {journal}
  {Physical Review B}\ }\textbf {\bibinfo {volume} {31}},\ \bibinfo {pages}
  {4966} (\bibinfo {year} {1985})}\BibitemShut {NoStop}%
\bibitem [{\citenamefont {Buffat}\ \emph {et~al.}(1986)\citenamefont {Buffat},
  \citenamefont {Chevalier}, \citenamefont {Tulier}, \citenamefont {Lloret},\
  and\ \citenamefont {Etourneau}}]{buffat}%
  \BibitemOpen
  \bibfield  {author} {\bibinfo {author} {\bibfnamefont {B.}~\bibnamefont
  {Buffat}}, \bibinfo {author} {\bibfnamefont {B.}~\bibnamefont {Chevalier}},
  \bibinfo {author} {\bibfnamefont {M.~H.}\ \bibnamefont {Tulier}}, \bibinfo
  {author} {\bibfnamefont {B.}~\bibnamefont {Lloret}}, \ and\ \bibinfo {author}
  {\bibfnamefont {J.}~\bibnamefont {Etourneau}},\ }\href@noop {} {\bibfield
  {journal} {\bibinfo  {journal} {Journal of Solid State Chemistry}\ }\textbf
  {\bibinfo {volume} {59}},\ \bibinfo {pages} {17} (\bibinfo {year}
  {1986})}\BibitemShut {NoStop}%
\bibitem [{\citenamefont {Vildosola}\ \emph {et~al.}(2004)\citenamefont
  {Vildosola}, \citenamefont {Llois},\ and\ \citenamefont
  {Sereni}}]{vildosola}%
  \BibitemOpen
  \bibfield  {author} {\bibinfo {author} {\bibfnamefont {V.}~\bibnamefont
  {Vildosola}}, \bibinfo {author} {\bibfnamefont {A.~M.}\ \bibnamefont
  {Llois}}, \ and\ \bibinfo {author} {\bibfnamefont {J.~G.}\ \bibnamefont
  {Sereni}},\ }\href@noop {} {\bibfield  {journal} {\bibinfo  {journal}
  {Physical Review B}\ }\textbf {\bibinfo {volume} {69}},\ \bibinfo {pages}
  {125116} (\bibinfo {year} {2004})}\BibitemShut {NoStop}%
\end{thebibliography}%

\end{document}